\documentclass[aps,prd,twocolumn,amsmath,amssymb,groupedaddress]{revtex4-2}
\usepackage{hyperref}
\hypersetup{
    colorlinks=true,
    linkcolor=blue,
    filecolor=blue,      
    urlcolor=blue,
    citecolor=blue,
}

\begin{document}


\title{Canonical Analysis of Brans-Dicke Theory Addresses Hamiltonian Inequivalence between Jordan and Einstein Frames }


\author{Gabriele Gionti, S.J. }
\email[]{ggionti@specola.va}
\affiliation{Specola Vaticana, V-00120 Vatican City, Vatican City State,
Vatican Observatory Research Group, Steward Observatory, The University Of Arizona,
933 North Cherry Avenue, Tucson, Arizona 85721, USA}
\affiliation{INFN, Laboratori Nazionali di Frascati, Via E. Fermi 40, 00044 Frascati, Italy.}


\date{\today}

\begin{abstract}
 Jordan and Einstein frames are studied under the light of Hamiltonian formalism. Dirac's constraint theory for  Hamiltonian systems is applied to Brans-Dicke theory in the Jordan Frame. In both Jordan and Einstein frames, Brans-Dicke theory has four secondary first class constraints and their constraint algebra is closed. We show, contrary to what is generally believed, the Weyl (conformal) transformation, between the two frames, is not a canonical transformation, in the sense of Hamiltonian formalism. This addresses quantum mechanical inequivalence as well. A canonical transformation is shown.  
 \end{abstract}

\maketitle

\section{\label{Jordan-Einstein}Introduction}

It is fairly well known that we never measure in physics absolute quantities, but ratios of absolute quantities. In fact we need to define a unit of measurement $u$ and determine how many times this unit of measurement is contained in the quantity we want to measure. For example suppose we work in natural units where the mass has the dimension of the inverse of length \cite{Dicke}.  Be $m_p$ the proton mass respect to unit of measurement ${m}_u$ and rescale the unit of measurement by a factor ${\lambda}^{-1}$, that is ${\tilde m}_u={\lambda}^{-1}{m}_u$, this implies that  in this new unit of measurement ${\tilde m}_p={\lambda}^{-1}{m}_p$ and the ratio \cite{Faraoni2006} stays constant 

\begin{equation}
\frac{{\tilde m}_p}{{\tilde m}_u}=\frac{{\lambda}^{-1}{m}_p}{{\lambda}^{-1}{m}_u}=\frac{m_p}{m_u}\,.
\label{physicalequiv}
\end{equation}

This rescaling appears more intuitive repeating these reasoning on length scales. In fact, in natural units,  \cite{Dicke} the above rescaling on the masses implies a length rescaling $d{\tilde x}^{\mu}={\lambda}dx^{\mu}$ and on the metric coefficients ${\tilde g}_{\mu\nu}={\lambda}^{2} g_{\mu\nu}$. Therefore \cite{Dicke} invariance of the physical observables under rescaling of units of measurements implies invariance under Weyl rescaling of the metric tensor. This is at the basis of the physical equivalence between Jordan and Einstein frame. 

Nowadays the general procedure \cite{Faraoni2006} is to start with a scalar-tensor theory action \cite{Dyer}  with the Gibbons-Hawking-York (GHY) boundary term \cite{gibbons&hawking} \cite{york1} \cite{york2} in what is called {\it Jordan frame}

\begin{eqnarray}
S&=&\int_{M}d^{n}x{\sqrt{-g}}\left(f(\phi)R-\frac{1}{2}\lambda(\phi)g^{\mu\nu}\partial_{\mu}\phi\partial_{\nu}\phi -U(\phi)\right) \nonumber \\
&+&2\int_{\partial M}d^{n-1}{\sqrt{h}}f(\phi)K \,.
\label{scalartensor}
\end{eqnarray}

where $f(\phi)$ is a generic function of $\phi$ as well as $\lambda(\phi)$, $K$ is the trace of the extrinsic curvature. This theory represents a generic scalar field non-minimally coupled to the gravitational field. If we perform the variation with respect the metric $g_{\mu\nu}(x)$ and set its variation $\delta g_{\mu\nu}(x)=0$ on the boundary,  we get the equations of motion for it 

\begin{equation}
f(\phi)\left(R_{\mu\nu}-\frac{1}{2}g_{\mu\nu}R\right)+g_{\mu\nu}\Box f(\phi)-\nabla_{\mu}\nabla_{\nu}f(\phi)=T^{\phi}_{\mu\nu}, 
\label{Einsteinequiv}
\end{equation}

where 
\begin{equation}
T^{\phi}_{\mu\nu}=\frac{1}{2}\left(\lambda(\phi)\partial_{\mu} \phi \partial _{\nu} \phi -\frac{1}{2}g_{\mu\nu}\lambda(\phi)g^{\alpha\beta}\partial_{\alpha} \phi \partial _{\beta} \phi \right)\,.
\label{tensorimpu}
\end{equation}

Variations respect to $\phi(x)$ and imposing these variations are zero at the boundary $\delta \phi(x)=0$ provide equation of motion for the scalar field $\phi (x)$

\begin{equation}
f'(\phi)R+\frac{1}{2}\lambda'(\phi)(\partial \phi)^{2}+\lambda(\phi) \Box \phi -U'(\phi)=0\;\;\;\,.
\label{eqaphi}
\end{equation}

In the literature one passes from the Jordan to the { \it Einstein frame} \cite{Dicke} \cite{Faraoni2006} through a Weyl transformation of the metric, above mentioned, which now, for convenience, we choose to be

\begin{equation}
{\tilde g}_{\mu\nu}=\Big(16\pi G f(\phi)\Big)^{\frac{2}{n-2}}g_{\mu\nu}\;, 
\label{Weyltrans}
\end{equation}

${\tilde g}_{\mu\nu}$ being the metric tensor in the Einstein frame. In the Einstein frame the action \eqref{scalartensor} becomes 

\begin{eqnarray}
S&=&\int_{M}d^{n}x{\sqrt{-{\tilde g}}}\left(\frac{1}{16\pi G}{\tilde R}-A(\phi){\tilde g}^{\mu\nu}\partial_{\mu}\phi\partial_{\nu}\phi -V(\phi)\right) \nonumber \\
&+&\frac{1}{8\pi G}\int_{\partial M}d^{n-1}{\sqrt{\tilde h}}{\tilde K}
\label{scalartensorEF},
\end{eqnarray}

where 

\begin{eqnarray}
A(\phi)&=&\frac{1}{16\pi G}\left(\frac{\lambda(\phi)}{2f(\phi)}+\frac{n-1}{n-2}\frac{(f'(\phi))^2}{f^2(\phi)}\right), \nonumber \\ 
V(\phi)&=&\frac{U(\phi)}{[16\pi G f(\phi)]^{\frac{n}{n-2}}}.
\label{AandV}
\end{eqnarray}

\noindent varying this equation respect to ${\tilde g}^{\mu\nu}$ we get Einstein Equations and varying respect to $\phi$ we get the equation for $\phi(x)$. As is well known \cite{Dicke} \cite{Faraoni2006}, if $(g_{\mu\nu}(x),\phi(x))$ is a solution of the equations in the Jordan frame, then, by construction, $(\tilde{g}_{\mu\nu}(x,\phi), \phi(x))$ is solution of the equations in the Einstein frame. Therefore the two frames are physically equivalent provided the scaling relations among observables quantities in the two frames \cite{Dicke} \cite{Faraoni2006} \cite{Cho1992}. Recently much work has been devoted to the study of Hamiltonian equivalence between the two frames \cite{Deruelle2009} \cite{Ezawa2009} as well as at quantum equivalence \cite{Falls2018} \cite{Kamenshchik2014} \cite{Ohta2017} \cite{Filippo2013}. In the following section we will perform the Dirac's constraint Hamiltonian analysis \cite{dirac1966} \cite{Esposito1992}(see also \cite{Olmo} \cite{Gielen} \cite{floreaniniJackiw} \cite{costagirotti} \cite{faddeevJackiw} for complementary cases) of Brans-Dicke theory and we will continue with the same analysis in the Einstein frame in order to confront and contrast these results in the two frames.

\section{\label{Hamiltogen}Hamiltonian analysis of Brans-Dicke theory}

Brans-Dicke theory \cite{Brans1961} is a particular case of \eqref{scalartensor} when $f(\phi)=\phi$ and $\lambda(\phi)=\frac{\omega}{\phi}$ \cite{Dyer}:

\begin{eqnarray}
S&=&\int_{M}d^{4}x\sqrt{-g}\left(\phi\;{}^{4}R-\frac{\omega}{\phi}g^{\mu\nu}\partial_{\mu}\phi \partial_{\nu} \phi -U(\phi)\right)\nonumber \\
&+& 2\int_{\partial M} d^3x \sqrt{h}\phi K\;\;\;\;.
\label{BDaction}
\end{eqnarray}

We implement the ADM-decomposition \cite{ADM}, the Space-Time $(M,g)$ is $M=R\times \Sigma$ \cite{Esposito1992}; $R$ is a one dimensional space, the time direction, $\Sigma$ is a three dimensional space-like surface embedded in $M$. The ADM metric tensor $g$ has the form as in  \cite{Esposito1992}. The ADM decomposition of the trace of the Ricci tensor ${}^{4} R$ is \cite{DeWitt1967} 
\begin{eqnarray}
&&{\sqrt{-g}}\; {}^{4}R=N\sqrt{h}\left( {}^{3}R+K_{ij}K^{ij}-K^2\right)-\\ \nonumber
&&\left(2K\sqrt{h}\right),_{0}+2f^{i}_{,i}\; ;\; f^{i}\equiv \sqrt{h}\left(KN^{i}-h^{ij}N_{,j}\right),
\end{eqnarray}

\noindent and involves terms which disappear, respectively, by the introduction of the boundary term in \eqref{BDaction} and assuming $\Sigma$'s boundary compact.   
$N=N(t,x)$ is the so called lapse function and $N^{i}=N^{i}(t,x)$ are the shift functions. The ADM Lagrangian density $\mathcal{L}_{ADM}$ is, 

\begin{eqnarray}\label{eq:Lagrangian2}
\mathcal{L}_{ADM}&=& {\sqrt{h}} \Bigg[N \phi\left( {}^{(3)}R+K_{ij}K^{ij}-K^2\right)\nonumber \\
&-&\frac{\omega}{N\phi}\left(N^2 h^{ij}D_i\phi D_j\phi- (\dot{\phi}-N^i D_i\phi)^2\right)  \\
&+&2K (\dot{\phi}-N^iD_i\phi )-NU(\phi)+2h^{ij}D_iN D_j\phi\Bigg]\;. \nonumber
\end{eqnarray}

The canonical momenta $(\pi, \pi_{i}, \pi^{ij}, \pi_{\phi})$ associated  to $(N, N^{i}, h_{ij}, \phi)$ are then

\begin{eqnarray}
\pi &=&\frac{\partial {\mathcal L}_{ADM}}{\partial \dot{N}}\approx 0 \ , \pi_i=\frac{\partial {\mathcal L}_{ADM}}{\partial \dot{N}^i}\approx 0 \ ,\pi^{ij}=\frac{\partial {\mathcal L}_{ADM} }{\partial \dot{h}_{ij}}    \nonumber \\
&=&-{\sqrt{h}}\left[ \phi \Big(K^{ij}-Kh^{ij}\Big)+\frac{h^{ij}}{N}\Big(\dot{\phi}-N^iD_i\phi\Big)\right] \label{pippo1}   \ , \ \\
\pi_\phi&=&\frac{\partial {\mathcal L}_{ADM}}{\partial \dot{\phi}}={\sqrt{h}}\left( 2K+\frac{2\omega}{N\phi}(\dot{\phi}-N^iD_i\phi)\right)\;, \nonumber
\end{eqnarray}

\noindent which show the momenta $\pi$ and $\pi_i$ associated to the lapse $N$ and shifts $N^i$ are primary constraints according to the theory of Dirac's constrained systems \cite{dirac1966} \cite{Esposito1992}. 
Once we have defined the Legendre transformation \eqref{pippo1} to pass from {\it velocities} to momenta, we are able to define the Hamiltonian density ${\mathcal {H}}_{ADM}$  knowing the Lagrangian density $\mathcal{L}_{ADM}$  

\begin{equation}
{\mathcal H}_{ADM}={\pi}^{ij}{\dot {h}}_{ij}+{\pi}_{\phi}{\dot \phi}-\mathcal{L}_{ADM}\;.
\label{hamiltodefin}
\end{equation}

This definition holds on the constraint surface defined by the Dirac's primary constraints $\pi \approx 0 $ and $\pi^{i}\approx 0 $ \cite{dirac1966} \cite{Esposito1992} found above \eqref{pippo1}.
Therefore the Hamiltonian density ${\mathcal H}_{ADM}$ is ($\pi_h\equiv \pi^{ij}h_{ij}$)

\begin{eqnarray}
&&{\mathcal{H}}_{ADM}={\sqrt{h}}\Bigg\{ N\left[-\phi\;  {}^{3}R+\frac{1}{\phi h}\left( \pi^{ij}\pi_{ij}-\frac{{\pi_h}^2}{2}\right)\right] \nonumber \\
&+& \frac{N\omega}{\phi}D_i\phi D^i\phi 
+N2D^iD_i\phi                                     
 +NV(\phi) \label{hamiltoeff1}  \\
 &+&\frac{1}{2 h\phi}\left(\frac{N}{3+2 \omega}\right)
 (\pi_h - \phi\pi_{\phi})^2\Bigg\}
 -2N^iD_j\pi^{j}_{i}+N^iD_i\phi \pi_{\phi}\;, \nonumber
\end{eqnarray}

and can be written in the following form 

\begin{equation}
{\mathcal{H}}_{ADM}=N{\mathcal H}+N^{i}{\mathcal H}_{i},
\label{scompositio}
\end{equation}

\noindent where the $\mathcal H$ is the Hamiltonian density constraint, and is just the quantity in square parenthesis of \eqref{hamiltoeff1} divided by $N$
and ${\mathcal {H}}_i$ is the momentum constraint 

\begin{equation}
{\mathcal {H}}_i= -2D_j\pi^{j}_{i}+D_i\phi \pi_{\phi}\;. 
\label{momentumcons}
\end{equation}

The total Hamiltonian $H_{T}$ \cite{Esposito1992} is at this point

\begin{equation}
H_{T}=\int d^{3}x \left(\lambda \pi + \lambda^{i}\pi_{i}+N{\mathcal{H}}+N^i{\mathcal{H}}_{i} \right)\;\,, 
\label{hamiltonianatot}
\end{equation}

\noindent where $\lambda=\lambda(t,x)$ and $\lambda^{i}(t,x)$ are Lagrange multipliers. If we indicate the canonical variables $(N,N^i, h_{ij}, \pi, \pi_{i}, \pi^{ij})$ generically with $(Q^i,\Pi_i)$ the Poisson Brackets between two arbitrary function $A$ and $B$ of the canonical variables is 

\begin{equation}
\left\{A,B\right\}=\int d^3y \left(\frac{\delta A}{\delta Q^{i}(y)}\frac{\delta B}{\delta \Pi_{i}(y)}-\frac{\delta A}{\delta \Pi_{i}(y)}\frac{\delta B}{\delta Q^{i}(y)}\right).
\label{PoissonBra}
\end{equation}

Following \cite{Menotti2017}, it is possible to show the momentum constraints ${\mathcal{H}}_i$ are the generators of the space-diffeomorphisms on the three-dimesional space-like surface $\Sigma$. The constraint algebra among the momentum constraints, and momentum constraints with Hamiltonian constraint can be easily calculated \cite{Menotti2017} and provides the same results as for Einstein Geometrodynamics \cite{Menotti2017}, \cite{Kuchar1}

In fact, imposing the primary constraint $\pi \approx 0$ and $\pi_i \approx 0$ be preserved on the dynamic, we get 

\begin{equation}
{\dot \pi}=\{\pi, H_T\}=-{\mathcal H} \approx 0 \, ,
\label{hamiltonianconstro}
\end{equation}

and 

\begin{equation}
{\dot {\pi}_{i}}=\{\pi_i, H_T\}=-{\mathcal H}_i \approx 0\, ,
\label{momentumconstro}
\end{equation}

We are now in the position to calculate the preservation of the secondary constraint along the dynamic. In doing this we will follow reference \cite{Menotti2017} adapted to our case of the Brans-Dicke theory. First we notice the following 

\begin{equation}
\left\{ h_{ij}(x), \int d^{3}y N^{l}(y){\mathcal {H}}_{l}(y) \right\}={\cal L}_{\mathbf N} h_{ij}(x)\,,
\label{suh}
\end{equation}

where ${\cal L}_{\mathbf N}$ is the Lie derivative along the three-dimensional vector field ${\mathbf N}$ defined by the shifts functions $N^l$. In an analogous way, but with a bit longer calculation \cite{Menotti2017}

\begin{equation}
\left\{ \pi^{ij}(x), \int d^{3}y N^{l}(y){\mathcal {H}}_{l}(y) \right\}={\cal L}_{\mathbf N} \pi^{ij}(x)\,,
\label{suh}
\end{equation}

We observe that 

\begin{eqnarray}
&&\left\{\phi(x), \int N^l(y) {\mathcal H}_l(y)d^3 y\right\}= \label{sufi} \\
&&\frac{\delta}{\delta \pi_{\phi}(x)} \int d^3 y \pi_{\phi}(y)D_i\phi(y)N^i (y) 
=N^i (x) D_i\phi(x)={\cal L}_{\mathbf N} \phi(x)\,, \label{sufi} \nonumber
\end{eqnarray}

while repeating the same reasoning on the momentum $\pi_{\phi}$ conjugated to $\phi$, we obtain 

\begin{eqnarray}
&&\left\{\pi_{\phi}(x), \int N^l(y){ {\mathcal H}_l}(y)d^3 y\right\}= \\
&&-\frac{\delta}{\delta \phi (x)} \int d^3 y \pi_{\phi}(y)D_i\phi(y)N^i (y) \nonumber
=D_i \left(\pi_{\phi}(x)N^i(x)\right)\, ,   
\label{supi}
\end{eqnarray}

The momenta calculated by the Legendre transformation using the Lagrangian ${\mathcal {L}}_{ADM}$ are densities as well, as it is immediate looking at \eqref{eq:Lagrangian2} 
\eqref{pippo1}. Then $\frac{\pi_{\phi}}{\sqrt h}$ is a scalar function. So

\begin{eqnarray}
{\cal L}_{\mathbf N} \pi_{\phi}(x)={\cal L}_{\mathbf N}\left(\sqrt{h}\left(\frac{\pi_{\phi}}{\sqrt h}\right)\right)\nonumber \\
={\cal L}_{\mathbf N}(\sqrt h)\frac{\pi_{\phi}}{\sqrt h}+{\sqrt h}{\cal L}_{\mathbf N}\left(\frac{\pi_{\phi}}{\sqrt h}\right)\nonumber\\
={\pi_{\phi}} D_i N^i+{\sqrt {h}}\partial_{i}\left(\frac{\pi_{\phi}}{\sqrt h}\right)N^i  \\
=\pi_{\phi}\frac{1}{\sqrt h}\partial_{i}({\sqrt h}N^i)+\partial_{i}(\pi_{\phi})N^i-\frac{\pi_{\phi}}{2 h}\partial_i (h)N^i\nonumber\\
=\pi_{\phi} \partial_i N^i+\partial_i (\pi_{\phi})N^i=D_i\left(\pi_{\phi}(x) N^i(x) \right)\,, \nonumber 
\label{tuttipasspi}
\end{eqnarray}

therefore $\int d^{3}x N^{l}{\mathcal {H}}_{l}$ is the generator of space diffeomorphisms on the three surfaces $\Sigma$ of the canonical variables $(h_{ij}, \phi, \pi^{ij}, \pi_{\phi})$, and any function $F(h_{ij}, \phi, \pi^{ij}, \pi_{\phi})$ of them, in particular of the density functions $\mathcal H$ and ${\mathcal H}_i$. Therefore we have (following \cite{Menotti2017})

\begin{equation}
{\cal L}_{\mathbf N} {\mathcal H}_i={\sqrt h}{\cal L}_{\mathbf N}\frac {{\mathcal H}_i}{\sqrt h}+ {{\mathcal H}_i}D_l N^l=N^l \partial_l {\mathcal H}_i + {\mathcal H}_l \partial_i N^l + {\mathcal H}_l \partial_i N^l \;,
\label{sumomentum}
\end{equation}
 
 which entitle us to write 
 
 \begin{equation}
 \left\{{\mathcal H}_i, \int N^s(y) {\mathcal H}_s(y)d^3 y\right\} =N^l \partial_l {\mathcal H}_i + {\mathcal H}_l \partial_i N^l + {\mathcal H}_l {\partial_i} N^l\;,
 \label{informa}
 \end{equation}
 
 and then the constraint algebra among the momenta constraints
 
 \begin{equation}
  \left\{{\mathcal H}_i(x), {\mathcal H}_j(x')\right\} = {\mathcal H}_i(x') \partial_j \delta(x,x')- {\mathcal H}_i(x) {\partial_j}' \delta(x,x')\;.  
  \label{commuto}
  \end{equation}
  
  As regard the Hamiltonian constraint ${\mathcal H}$, we start from the following 
  
  \begin{equation}
  {\cal L}_{\mathbf N}{\mathcal H}= \left\{{\mathcal H}, \int N^s(y) {\mathcal H}_s(y)d^3 y\right\}\;, 
  \label{liehamilt}
  \end{equation}
  
  and repeating the same reasoning above, we get 
  
  \begin{equation}
 {\cal L}_{\mathbf N} {\mathcal H}={\sqrt h}{\cal L}_{\mathbf N}\frac {\mathcal H}{\sqrt h}+\frac {\mathcal H}{\sqrt h} {\cal L}_{\mathbf N}{\sqrt h} =N^l \partial_l {\mathcal H} + {\mathcal H} \partial_i N^i \;, 
 \label{rifa}
  \end{equation}  
  
 finally we can write

\begin{equation}
 \left\{{\mathcal H}(x), {\mathcal H}_i(x')\right\}=-{\mathcal H}(x'){{\partial}'_i}\delta(x,x')\,.
 \label{harifa}
 \end{equation}

As usual, the calculation of the Poisson brackets of Hamiltonian constraint is more involved. 

\begin{equation}
\left\{\int d^{3}x N(x){\cal{H}}(x),\int d^{3}x' N'(x') {\cal{H}}(x')\right\}\,.
\label{hamiltonian-}
\end{equation}

Following \cite{Menotti2017}, we observe that $h_{ij}$ is present in ${\cal{H}}$ both in algebraic and non algebraic way via its derivaties. Its conjugate variable $\pi^{ij}$ only in algebraic way. The same thing  happens for $\phi$ and $\pi_{\phi}$. The algebraic-algebraic variation contributes to the Poisson brackets \eqref{hamiltonian-}, as regard the couple $(h_{ij}, \pi^{ij})$, always zero as shown in \cite{Menotti2017}, which we report here for clarity  (the same holds for $(\phi, \pi_{\phi})$ ),   

\begin{equation}
\frac{\delta {\cal{H}}(x)}{\delta h_{ij}(y)}=f_{ h_{ij}}(x)\delta(x,y), \,\,\,\,\,\,\,\frac{\delta {\cal{H}}(x)}{\delta \pi^{ij}(y)}=f_{ \pi ^{ij}}(x)\delta(x,y),
\label{func}
\end{equation}

and then 

\begin{eqnarray}
&&\int d^{3}y\Big(f_{ h_{ij}}(x)\delta(x,y)f_{ \pi^{ij}}(x')\delta(x',y)N(x)N'(x') \nonumber \\
&&-f_{ \pi ^{ij}}(x)\delta(x,y)f_{ h_{ij}}(x')\delta(x',y)N(x)N'(x')\Big)=0\,. 
\label{pedante}
\end{eqnarray}

The only non-zero terms of the Poisson brackets originate by non-algebraic variation of $h_{ij}$ combined with the algebraic variation of $\pi^{ij}$ and by the non algebraic variation of $\phi$ with the algebraic variation of $\pi_{\phi}$. The non algebraic contribution to the Poisson brackets of the variation $\delta h_{ij}$ of $h_{ij}$ is contained both in the term $-N\sqrt{h} \phi\;  {}^{3}R$, which can be derived, easily, from reference \cite{Menotti2017} 

\begin{equation}
-\int d^{3}y\sqrt{h}\delta h_{ij}\left((D^{i}D^{j})(N\phi)-h^{ij}(D^{k}D_{k})(N\phi)\right)\, ,
\label{varioR}
\end{equation}

and in the term $2 N\sqrt{h} D_i D^i\phi$ (integrated by parts is equivalent to $-2 \sqrt{h} D_i ND^{i}\phi$), that is  

\begin{equation} 
-2\int d^3y\sqrt{h}\delta h_{ij}\left(\frac{1}{2}h^{ij}(D^k N)(D_k \phi)-(D^{i}N)(D^{j}\phi)\right) \, .
\label{doublederiv}
\end{equation}

In a parallel manner, the non algebraic contribution to the Poisson brackets of the variation $\delta \phi$ of $\phi$ is  contained in the terms $-N\sqrt{h}\frac{\omega}{\phi}D_i\phi D^i\phi$ and $-2N\sqrt{h}D_i D^i\phi$, that is 

\begin{equation}
\int d^{3}y\sqrt{h} \delta\phi \left(-\frac{2\omega}{\phi}(D_{k}N)(D^{k}\phi) + 2(D^k D_k)(N)\right)\, .
\label{varphiHam}
\end{equation}

At this point, it is immediate to write

\begin{eqnarray}
&&\int d^{3}y \frac{\delta \int d^{3}x N(x){\cal{H}}(x)}{\delta h_{ij}(y)} \frac{\delta \int d^{3}x' N'(x'){\cal{H}}(x')}{\delta \pi^{ij}(y)} \nonumber \\
&&=-\Bigg(\int d^{3}y\left[(D^{i}D^{j})(N\phi)-h^{ij}(D^{k}D_{k})(N\phi)\right] \times \nonumber \\ 
&&\frac{N'}{\phi}\left(\pi_{ij}-h_{ij}\frac{(2+2\omega)\pi_h + \phi \pi_{\phi}}{(2\omega +3)} \right) \Bigg) \\ \nonumber 
&&-2\int d^3y\left(\frac{1}{2}h^{ij}(D^k N)(D_k \phi)-(D^{i}N)(D^{j}\phi)\right) \times \\ \nonumber
&&\frac{N'}{\phi} \left(\pi_{ij}-h_{ij}\frac{(2+2\omega)\pi_h + \phi \pi_{\phi}}{(2\omega +3)} \right)\, ,
\label{intermedia1}
\end{eqnarray}

and

\begin{eqnarray}
&&\int d^{3}y \frac{\delta \int d^{3}x N(x){\cal{H}}(x)}{\delta \phi(y)} \frac{\delta \int d^{3}x' N'(x'){\cal{H}}(x')}{\delta \pi_{\phi}(y)} \nonumber \\
&&=\int d^{3}y\left(-\frac{2\omega}{\phi}(D_{k}N)(D^{k}\phi) + 2(D^k D_k)(N)\right)\times  \nonumber \\
&&N' \left(-\frac{(\pi_{h}-\phi\pi_{\phi})}{(2\omega + 3)}\right).
\label{intermedia2}
\end{eqnarray}

Grouping together all the terms, we get

\begin{eqnarray}
&&\left\{\int d^{3}x N(x){\cal{H}}(x),\int d^{3}x' N'(x') {\cal{H}}(x')\right\} \nonumber = \\
&&-\Bigg(\int d^{3}y\left[(D^{i}D^{j})(N\phi)-h^{ij}(D^{k}D_{k})(N\phi)\right]\frac{N'}{\phi} \times\\ \nonumber
&&\left(\pi_{ij}-h_{ij}\frac{(2+2\omega)\pi_h + \phi \pi_{\phi}}{(2\omega +3)} \right)-N\mapsto N' \Bigg) \\ \nonumber 
&&+\int d^{3}y\left(-\frac{2\omega}{\phi}(D_{k}N)(D^{k}\phi) + 2(D^k D_k)(N)\right)N' \times \\ \nonumber
&&\left(-\frac{(\pi_{h}-\phi\pi_{\phi})}{(2\omega + 3)}\right)-N\mapsto N' \\ \nonumber
&&-2\int d^3y\left(\frac{1}{2}h^{ij}(D^k N)(D_k \phi)-(D^{i}N)(D^{j}\phi)\right)\frac{N'}{\phi} \times \\ \nonumber
&&\left(\pi_{ij}-h_{ij}\frac{(2+2\omega)\pi_h + \phi \pi_{\phi}}{(2\omega +3)} \right) -N\mapsto N' \label{longa}.
\end{eqnarray}

Simplifying, the second member of the previous equation becomes

\begin{eqnarray}
&&-\Bigg(\int d^{3}y \left(D^i D^j \right)(N) N' (2\pi_{ij}) -N\mapsto N' \Bigg)  + \nonumber \\
+&&\int d^{3}y \left(D^i D_i \right)(N) N' \bigg(\frac{(2+2\omega)\pi_{h}+\phi \pi_\phi}{2 \omega + 3} -  \nonumber \\
&&-\frac{2\omega\pi_{h}+3\phi\pi_{\phi}}{2\omega + 3}-2\frac{\pi_{h}-\phi\pi_{\phi}}{2\omega+3} \bigg)-N\mapsto N' \nonumber \\
&&+\int d^{3}y (D_i N)(D^i \phi)\frac{N'}{\phi}\Bigg(\frac{2\omega(\pi_h -\phi \pi_{\phi})}{2\omega+3} +\nonumber \\
&&+\frac{2\omega\pi_{h}+3\phi\pi_{\phi}}{2\omega+3} - 2\frac{2\omega\pi_h +3\phi \pi_{\phi}}{2\omega+3} \Bigg)-N\mapsto N' \, . \label{menolonga}
\end{eqnarray}

And then we have

\begin{eqnarray}
&&\left\{\int d^{3}x N(x){\cal{H}}(x),\int d^{3}x' N'(x') {\cal{H}}(x')\right\} \\ \nonumber 
&&=\int d^{3}y \left(ND^{i}(N')-N'D^{i}(N)\right){\mathcal H}_{i}\;, 
\label{hamil-hamil}
\end{eqnarray}

equivalent to 

\begin{equation}
\{{\cal{H}}(x),{\cal{H}}(x')\}={\cal {H}}^{i}(x)\partial_{i}\delta(x,x')-{\cal H}^{i}(x'){\partial}'_{i}\delta(x,x') .
\label{differential} 
\end{equation}

Interesting enough it is the same secondary first class constraint algebra like pure Einstein Geometrodynamics. As extensively argued in \cite{Kuchar2} and \cite{Kuchar1}, once matter source is introduced with its own canonical variables, many different inequivalent theory of gravity coupled with matter can generate the same constraint algebra \eqref{commuto} \eqref{harifa} \eqref{differential}.

\section{\label{comparison}Canonical Transformations }

One notice, see for example \cite{Garay1992} \cite{Deruelle2009}, the transformation \eqref{Weyltrans} entails ADM metric in the Einstein frame 
\begin{eqnarray}
\tilde {g}&=&-({\tilde N}^{2}-{\tilde N}_{i}\tilde{N}^{i})dt \otimes dt \nonumber \\
&+&\tilde{N}_{i}(dx^{i} \otimes dt
+dt \otimes dx^{i})+\tilde{h}_{ij}dx^{i} \otimes dx^{j}\;\;, 
\label{EFmetricADM}
\end{eqnarray}

\noindent where

\begin{eqnarray}
\tilde{N}&=&\left(16\pi G f(\phi) \right)^{\frac{1}{n-2}}N; \tilde{N}_i=\left(16\pi G f(\phi) \right)^{\frac{2}{n-2}}N_i;  \nonumber \\
\tilde{h}_{ij}&=&\left(16\pi G f(\phi) \right)^{\frac{2}{n-2}}h_{ij}.
\label{tilderelation}
\end{eqnarray}

 The scalar-tensor action \eqref{scalartensorEF} in the Einstein frame for $n=4$ for Brans-Dicke in the particular case $\lambda(\phi)=\frac{2\omega}{\phi}$ and $f(\phi)=\phi$  reduces to equation \eqref{scalartensorEF} with $A(\phi)=\frac{\left(\omega +\frac{3}{2}\right)}{16\pi G \phi^{2}}$.  

Applying the ADM decomposition in the Einstein Frame \eqref{EFmetricADM}, we can derive the ADM Lagrangian density 
${\mathcal{\tilde {L}_{\rm ADM}}}$. The canonical momenta in the Einstein Frame are defined through the Legendre transformation. Using Weyl (conformal) transformations \cite{Dabrowski2008}, one can confront them with the analogous quantities in the Jordan Frame   

\begin{eqnarray}
&&{\tilde \pi}^{ij}= \frac{\partial {\mathcal {\tilde L}}_{ADM} }{\partial \dot{\tilde{h}}_{ij}}=
-\frac{\sqrt{\tilde {h}}}{{16 \pi G}}\left( {\tilde K}^{ij}-{\tilde K}{\tilde h}^{ij}\right)=\frac{{\pi}^{ij}}{16\pi G\phi} \, ,\nonumber \\
&&{\tilde \pi}_\phi=\frac{\partial {\mathcal {\tilde L}}_{ADM}}{\partial \dot{\phi}}=\frac{\sqrt{\tilde {h}}(\omega +\frac{3}{2})}{8\pi G {\tilde N}{\phi}^2}\left(\dot{\phi}-{\tilde N}^i\partial_i\phi \right)\nonumber \\
&&=\frac{1}{\phi}(\phi \pi_{\phi}-\pi_{h}).
\label{momentumEF}
\end{eqnarray}

\noindent  The canonical Hamiltonian density ${\mathcal{H}}_{ADM}$ is defined in analogy to \eqref{hamiltodefin} $\left(\textrm{we pose}\; {\tilde V}(\phi)=16\pi G V(\phi)\right)$  

\begin{eqnarray}
&&{\mathcal{H}}_{ADM}=\frac{\sqrt{{\tilde h}}\tilde{N}}{16\pi G}\Bigg[ -{}^{3}{\tilde R}+\frac{(16\pi G)^2}{\tilde h}\left( {\tilde \pi}^{ij}{\tilde \pi}_{ij}-\frac{{{\tilde \pi}_h}^2}{2}\right) \nonumber \\
&&+ \frac{(\omega +\frac{3}{2})}{{\phi}^2}\partial_i\phi\partial ^i\phi \
+\frac{64(\pi G)^{2}{\phi}^{2}}{h(\omega + \frac{3}{2})} {\tilde \pi}_{\phi}^{2} +{\tilde V}(\phi)\Bigg] \\
 &&-2\tilde{N}^i{\tilde D}_j{\tilde \pi}^{j}_{i}+\tilde{N}^i\partial_i\phi {\tilde \pi}_{\phi}\;. \nonumber
\label{hamiltoeff}
\end{eqnarray}

The Hamiltonian constraint ${\mathcal{H}}$ is, in parallel to the previous section, the quantity in square parenthesis divided by $\tilde{N}$
while the momentum constraints ${\mathcal{H}}_{i}=-2{\tilde D}_j{\tilde \pi}^{j}_{i}+\partial_i\phi {\tilde \pi}_{\phi}$.  

The Hamiltonian ${\mathcal{H}}$ and momentum ${\mathcal{H}}_{i}$ constraints are first class constraints and behave like in standard Einstein geometrodynamics \cite{Kuchar1}.

We recall that in the Hamiltonian theory the transformation $(Q^i(q,p), P_i(q,p))$ between two sets of variables $(q^i,p_i)$ and $(Q^i,P_i)$  is canonical if the \textquotedblleft symplectic two form\textquotedblright $\omega=dq^i \wedge dp_i$ is invariant that is $\omega=dQ^i \wedge dP_i$, which is equivalent to say that the Poisson brackets fulfill the following conditions

\begin{eqnarray}
\{Q^{i}(q,p),P_{j}(q,p)\}_{q,p}&=&\delta^{i}_{j} \label{symplectic}  \\
\{Q^{i}(q,p),Q^{j}(q,p)\}_{q,p}&=&\{P_{i}(q,p),P_{j}(q,p)\}_{q,p}=0 .
\nonumber
\end{eqnarray}

The transformations \eqref{tilderelation} \eqref{momentumEF}, (cfr. \cite{Deruelle2009}) , represent a canonical, in Hamiltonian sense, change of variable. But if we include the lapse $N$ and shifts $N^i$, and their conjugate momenta $\pi$ and $\pi_{i}$, this is not completely true (cfr. \cite{Kiefer2017}) since 

\begin{equation}
\{{\tilde N},{\tilde \pi}_{\phi} \}=\frac{8\pi GN}{{\sqrt{16 \pi G\phi}}} \neq 0, \textrm{and}\; 
\{{\tilde N}_i,{\tilde \pi}_{\phi} \}=16 \pi G N_i \neq 0\, ,
\label{noncanonicalcond}
\end{equation}

where, obviously, the Poisson brackets are calculated in the Jordan Frame. Therefore, strictly speaking, it is not correct to pass form the Jordan Frame to the Einstein frame in oder to perform the constraint analysis of the Brans-Dicke theory as it is usually done (cfr. \cite{Garay1992}). The right Hamiltonian canonical transformations hold the lapse and the shifts ${\tilde {N^{*}}}=N$ and ${\tilde {N^{*}}}^{i}=N^{i}$ while $h_{ij}$ and $\phi$ and their respective momenta $\pi^{ij}$ and $\pi_{\phi}$ transform according to the equations \eqref{tilderelation} and \eqref{momentumEF}. These transformations generate an Anti-Newtonian Gravity as explained in \cite{Niedermaier2020}.  The ADM Hamiltonian ${\mathcal{H}}_{ADM}$, in this new set of variables, is

\begin{eqnarray}
&&{\mathcal{H}}_{ADM}=\frac{\sqrt{{\tilde h}}\tilde{N^{*}}(\phi)^{1\over 2}}{(16\pi G)^{1 \over 2}}\Bigg[ -{}^{3}{\tilde R}+\frac{(16\pi G)^2}{\tilde h}\left( {\tilde \pi}^{ij}{\tilde \pi}_{ij}-\frac{{{\tilde \pi}_h}^2}{2}\right) \nonumber \\
&&+ \frac{(\omega +\frac{3}{2})}{{\phi}^2}\partial_i\phi\partial ^i\phi \
+\frac{64(\pi G)^{2}{\phi}^{2}}{h(\omega + \frac{3}{2})} {\tilde \pi}_{\phi}^{2} +{\tilde V}(\phi)\Bigg]  \nonumber \\
 &&-2\tilde{N^{*}}^i{\tilde D}_j{\tilde \pi}^{j}_{i}+\tilde{N^{*}}^i\partial_i\phi {\tilde \pi}_{\phi}\;. 
\label{hamilteff}
\end{eqnarray}

This theory is canonically equivalent to Brans-Dicke but not equivalent to the Hamiltonian of Brans-Dicke theory in the Einstein Frame \eqref{hamiltoeff}. 

As a simple application, one can consider the high symmetrical case of flat FLRW universe with spatial curvature $k=0$

\begin{equation}
ds^{2}=-N^{2}(t)+a^{2}(t)\left(dr^2 + r^2d{\theta}^2 + r^2 sin^{2}\theta d{\varphi}^{2}\right).
\label{FLRWpiatta}
\end{equation}

We derive the Lagrangian function ${\mathcal{L}}$  of this mini-superspace model by substituting this metric in the action \eqref{BDaction} (cfr. \cite{Bonanno2017})

\begin{equation}
{\mathcal{L}}=-\frac{6a{\dot{a}}^2}{N(t)}\phi(t)-\frac{6a^2{\dot{a}}}{N(t)}{\dot{\phi}(t)}+\frac{\omega a^3}{N\phi(t)}(\dot{\phi}(t))^2-Na^3U(\phi(t))\,. 
\label{lagra}
\end{equation}

The \textquotedblleft configuration\textquotedblright variables are the Lapse $N=N(t)$, the scale factor of the Universe $a(t)$ of FLRW metric and the field $\phi=\phi(t)$ which now depends only on time $t$ for symmetry reasons. Given the mini-superspace Lagrangian $\mathcal{L}$, we can define the Hamiltonian ${\mathcal{H}}_{ADM}$ in analogy to \eqref{hamiltodefin}. We start with the definition of the canonical momenta

\begin{eqnarray}
\pi &=&\frac{\partial {\mathcal L}}{\partial \dot{N}}\approx 0 \ ,\pi_{a}=\frac{\partial {\mathcal L}}{\partial \dot{a}} =-\frac{12a{\dot{a}}}{N(t)}\phi(t)-\frac{6a^2{\dot{a}}}{N(t)}{\dot{\phi}(t)}   \ , \nonumber \\
\pi_\phi&=&\frac{\partial {\mathcal L}}{\partial \dot{\phi}}=-\frac{6a^2{\dot{a}}}{N(t)}+\frac{2\omega a^3}{N\phi(t)}\dot{\phi}(t)\,, 
\label{pippo2}
\end{eqnarray}

and 

\begin{eqnarray}
&&{\mathcal{H}}_{ADM}=N\bigg(-\frac{\omega{\pi}^{2}_{a}}{12a\phi (2\omega +3)}-\frac{ {\pi}_{a} \pi_{\phi}}{2 a^2 (2\omega +3)} \nonumber \\
&&\frac{ \phi {\pi}^2_{\phi}}{2 a^3 (2\omega +3)} + a^3 U(\phi) \bigg) \ ,
\label{hamiltopunto}
\end{eqnarray}

where the Hamiltonian constraint $\mathcal{H}$ is just the quantity under parenthesis. One can check that the following set of transformations 

\begin{eqnarray}
&&{{\tilde {N}}^{*}}=N \, \, ; \, {\tilde{\pi}}^{*}=\pi  \,\, ; \, \tilde{a}=(16\pi G\phi)^{1\over 2}a \, \, ; \nonumber \\
&&{\tilde \pi}_{a}=\frac{{\pi}_{a}}{16\pi G\phi} \, \,;\, \phi=\phi \, \, ; \, {\tilde \pi}_\phi=\frac{1}{\phi} \, ( \phi \pi_{\phi}-\frac{1}{2}a\pi_{a})\,,
\label{trasffino}
\end{eqnarray}

are canonical according to the definition \eqref{symplectic}. The ADM-Hamiltonian in this new canonical variables is 

\begin{eqnarray}
&&{\mathcal{H}}_{ADM}=\frac{{\tilde{a}}^{3}{\tilde{{N}}^{*}} {\phi}^{1\over 2}}{(16\pi G)^{1\over 2}}\bigg(-\frac{(16\pi G)^{3} \phi {\tilde \pi}_{a}}{24 {\tilde{a}}^4}+ \nonumber \\
&&+\frac{(16 \pi G)^{2} {\phi}^{2} {\pi}^{2}_{\phi}}{2 {\tilde a}^6 (2\omega +3)}+ {\tilde V}(\phi) \bigg) \, .
\label{hamiltopuntotrans}
\end{eqnarray}

This mini-superspace Hamiltonian has an overall multiplicative factor $(\phi)^{1\over 2}$,as the Hamiltonian constraint  in \eqref{hamilteff}. 
The set of transformations which differs from previous one for 
$\tilde{N}=(16\pi G\phi)^{1\over 2}N$ and ${\tilde{\pi}}=\frac{\pi}{(16\pi G\phi)^{1\over 2}}$, analogous to the transformation from the Jordan to the Einstein Frame \eqref{tilderelation} \eqref{momentumEF}, is not canonical for the same reason as in \eqref{noncanonicalcond}. 

If one wants to get rid of the variable $N$, and $\pi$, a method is to perform a gauge fixing. For example, one fixes $N=1$ (cfr. \cite{Bonanno2017}) and  treats it as a secondary Dirac's constraint. Performing Dirac's constraint analysis \cite{dirac1966}, $N$ and $\pi$ are eliminated by imposing strongly the secondary constraints and substituting the Poisson's brackets with the Dirac's brackets.

\section{{\label{conclusions}} Conclusions}

We have introduced the action of a scalar tensor theory of gravity with boundary terms and derived the equations of motion both in the Jordan and Einstein frame. Motivated by recent works on the quantum inequivalence between Jordan and Einstein Frame \cite{Falls2018} \cite{Ohta2017} \cite{Kamenshchik2014}, we have performed ADM-Dirac's constraint analysis of Brans-Dicke theory as a particular case of scalar tensor theory. This Hamiltonian analysis in the Jordan frame  exhibits secondary first class constraints, ${\mathcal H}$ and ${\mathcal H}_{i}$, whose Poisson brackets close like Einstein geometrodynamics \cite{Kuchar1}, although the calculation of the Poisson Brackets between Hamiltonian-Hamiltonian constraints is more involved. The Weyl (conformal) transformations from the Jordan to the Einstein frame result to be not a Hamiltonian canonical transformation. Therefore, strictly speaking, the procedure of making the constraint analysis of the Brans-Dicke theory by passing from Jordan to Einstein frame is not correct. Furthermore we are not sure that solutions of the equations of motions, in the Hamiltonian formalism, in the Jordan Frame are solution of the equations of motion also in the Einstein frame.  The Legendre map is not a one to one transformation for Dirac's constrained systems. This is one of the reasons of Hamiltonian in-equivalence between Jordan and Einstein frame. We exhibit a canonical transformation from  Brans-Dicke to a theory of gravity with non-minimally coupled matter, Brans-Dicke like. This fact addresses quantum inequivalence as well. In reference \cite{Banerjee2016}, it is shown Hamiltonian quantitation in the minusuperspace case with flat FLRW metric generates two physical inequivalent solutions in the two frames. As regards the path integral \cite{Falls2018} \cite{Kamenshchik2014} \cite{Filippo2013}, it has been already mentioned in-equivalence between the two frames. At level of pure speculations, following Dicke's reasoning of the physical equivalence between the two frames, it could be possible, in order to restore full physical equivalence also at Hamiltonian level, to pursue the path of a re-definition of the Poisson brackets like in non-commutative geometry \cite{Ezawa2009}.  

\begin{acknowledgments}
We thank A. Bonanno for encouraging us along the different stages of this work and for hospitality at OACT in Catania. We thank G. Esposito, N. Deruelle, M. Galaverni, F. Nesti, A. Kamenshchik, M. Reuter for useful discussions.
\end{acknowledgments}


\begin{thebibliography}{32}%
\makeatletter
\providecommand \@ifxundefined [1]{%
 \@ifx{#1\undefined}
}%
\providecommand \@ifnum [1]{%
 \ifnum #1\expandafter \@firstoftwo
 \else \expandafter \@secondoftwo
 \fi
}%
\providecommand \@ifx [1]{%
 \ifx #1\expandafter \@firstoftwo
 \else \expandafter \@secondoftwo
 \fi
}%
\providecommand \natexlab [1]{#1}%
\providecommand \enquote  [1]{``#1''}%
\providecommand \bibnamefont  [1]{#1}%
\providecommand \bibfnamefont [1]{#1}%
\providecommand \citenamefont [1]{#1}%
\providecommand \href@noop [0]{\@secondoftwo}%
\providecommand \href [0]{\begingroup \@sanitize@url \@href}%
\providecommand \@href[1]{\@@startlink{#1}\@@href}%
\providecommand \@@href[1]{\endgroup#1\@@endlink}%
\providecommand \@sanitize@url [0]{\catcode `\\12\catcode `\$12\catcode
  `\&12\catcode `\#12\catcode `\^12\catcode `\_12\catcode `\%12\relax}%
\providecommand \@@startlink[1]{}%
\providecommand \@@endlink[0]{}%
\providecommand \url  [0]{\begingroup\@sanitize@url \@url }%
\providecommand \@url [1]{\endgroup\@href {#1}{\urlprefix }}%
\providecommand \urlprefix  [0]{URL }%
\providecommand \Eprint [0]{\href }%
\providecommand \doibase [0]{https://doi.org/}%
\providecommand \selectlanguage [0]{\@gobble}%
\providecommand \bibinfo  [0]{\@secondoftwo}%
\providecommand \bibfield  [0]{\@secondoftwo}%
\providecommand \translation [1]{[#1]}%
\providecommand \BibitemOpen [0]{}%
\providecommand \bibitemStop [0]{}%
\providecommand \bibitemNoStop [0]{.\EOS\space}%
\providecommand \EOS [0]{\spacefactor3000\relax}%
\providecommand \BibitemShut  [1]{\csname bibitem#1\endcsname}%
\let\auto@bib@innerbib\@empty
\bibitem [{\citenamefont {Dicke}(1962)}]{Dicke}%
  \BibitemOpen
  \bibfield  {author} {\bibinfo {author} {\bibfnamefont {R.~H.}\ \bibnamefont
  {Dicke}},\ }\bibfield  {title} {\bibinfo {title} {{Mach's principle and
  invariance under transformation of units}},\ }\href
  {https://doi.org/10.1103/PhysRev.125.2163} {\bibfield  {journal} {\bibinfo
  {journal} {Phys. Rev.}\ }\textbf {\bibinfo {volume} {125}},\ \bibinfo {pages}
  {2163} (\bibinfo {year} {1962})}\BibitemShut {NoStop}%
\bibitem [{\citenamefont {Faraoni}\ and\ \citenamefont
  {Nadeau}(2007)}]{Faraoni2006}%
  \BibitemOpen
  \bibfield  {author} {\bibinfo {author} {\bibfnamefont {V.}~\bibnamefont
  {Faraoni}}\ and\ \bibinfo {author} {\bibfnamefont {S.}~\bibnamefont
  {Nadeau}},\ }\bibfield  {title} {\bibinfo {title} {{The (pseudo)issue of the
  conformal frame revisited}},\ }\href
  {https://doi.org/10.1103/PhysRevD.75.023501} {\bibfield  {journal} {\bibinfo
  {journal} {Phys. Rev.}\ }\textbf {\bibinfo {volume} {D75}},\ \bibinfo {pages}
  {023501} (\bibinfo {year} {2007})},\ \Eprint
  {https://arxiv.org/abs/gr-qc/0612075} {arXiv:gr-qc/0612075 [gr-qc]}
  \BibitemShut {NoStop}%
\bibitem [{\citenamefont {Dyer}\ and\ \citenamefont
  {Hinterbichler}(2009)}]{Dyer}%
  \BibitemOpen
  \bibfield  {author} {\bibinfo {author} {\bibfnamefont {E.}~\bibnamefont
  {Dyer}}\ and\ \bibinfo {author} {\bibfnamefont {K.}~\bibnamefont
  {Hinterbichler}},\ }\bibfield  {title} {\bibinfo {title} {{Boundary Terms,
  Variational Principles and Higher Derivative Modified Gravity}},\ }\href
  {https://doi.org/10.1103/PhysRevD.79.024028} {\bibfield  {journal} {\bibinfo
  {journal} {Phys. Rev.}\ }\textbf {\bibinfo {volume} {D79}},\ \bibinfo {pages}
  {024028} (\bibinfo {year} {2009})},\ \Eprint
  {https://arxiv.org/abs/0809.4033} {arXiv:0809.4033 [gr-qc]} \BibitemShut
  {NoStop}%
\bibitem [{\citenamefont {Gibbons}\ and\ \citenamefont
  {Hawking}(1977)}]{gibbons&hawking}%
  \BibitemOpen
  \bibfield  {author} {\bibinfo {author} {\bibfnamefont {G.~W.}\ \bibnamefont
  {Gibbons}}\ and\ \bibinfo {author} {\bibfnamefont {S.~W.}\ \bibnamefont
  {Hawking}},\ }\bibfield  {title} {\bibinfo {title} {Action integrals and
  partition functions in quantum gravity},\ }\href
  {https://doi.org/10.1103/PhysRevD.15.2752} {\bibfield  {journal} {\bibinfo
  {journal} {Phys. Rev. D}\ }\textbf {\bibinfo {volume} {15}},\ \bibinfo
  {pages} {2752} (\bibinfo {year} {1977})}\BibitemShut {NoStop}%
\bibitem [{\citenamefont {York}(1972)}]{york1}%
  \BibitemOpen
  \bibfield  {author} {\bibinfo {author} {\bibfnamefont {J.~W.}\ \bibnamefont
  {York}, \bibfnamefont {Jr.}},\ }\bibfield  {title} {\bibinfo {title} {{Role
  of conformal three geometry in the dynamics of gravitation}},\ }\href
  {https://doi.org/10.1103/PhysRevLett.28.1082} {\bibfield  {journal} {\bibinfo
   {journal} {Phys. Rev. Lett.}\ }\textbf {\bibinfo {volume} {28}},\ \bibinfo
  {pages} {1082} (\bibinfo {year} {1972})}\BibitemShut {NoStop}%
\bibitem [{\citenamefont {York}(1986)}]{york2}%
  \BibitemOpen
  \bibfield  {author} {\bibinfo {author} {\bibfnamefont {J.}~\bibnamefont
  {York}},\ }\bibfield  {title} {\bibinfo {title} {{Boundary terms in the
  action principles of general relativity}},\ }\href
  {https://doi.org/10.1007/BF01889475} {\bibfield  {journal} {\bibinfo
  {journal} {Found. Phys.}\ }\textbf {\bibinfo {volume} {16}},\ \bibinfo
  {pages} {249} (\bibinfo {year} {1986})}\BibitemShut {NoStop}%
\bibitem [{\citenamefont {Cho}(1992)}]{Cho1992}%
  \BibitemOpen
  \bibfield  {author} {\bibinfo {author} {\bibfnamefont {Y.}~\bibnamefont
  {Cho}},\ }\bibfield  {title} {\bibinfo {title} {{Reinterpretation of
  Jordan-Brans-Dicke theory and Kaluza-Klein cosmology}},\ }\href
  {https://doi.org/10.1103/PhysRevLett.68.3133} {\bibfield  {journal} {\bibinfo
   {journal} {Phys. Rev. Lett.}\ }\textbf {\bibinfo {volume} {68}},\ \bibinfo
  {pages} {3133} (\bibinfo {year} {1992})}\BibitemShut {NoStop}%
\bibitem [{\citenamefont {Deruelle}\ \emph {et~al.}(2009)\citenamefont
  {Deruelle}, \citenamefont {Sendouda},\ and\ \citenamefont
  {Youssef}}]{Deruelle2009}%
  \BibitemOpen
  \bibfield  {author} {\bibinfo {author} {\bibfnamefont {N.}~\bibnamefont
  {Deruelle}}, \bibinfo {author} {\bibfnamefont {Y.}~\bibnamefont {Sendouda}},\
  and\ \bibinfo {author} {\bibfnamefont {A.}~\bibnamefont {Youssef}},\
  }\bibfield  {title} {\bibinfo {title} {{Various Hamiltonian formulations of
  f(R) gravity and their canonical relationships}},\ }\href
  {https://doi.org/10.1103/PhysRevD.80.084032} {\bibfield  {journal} {\bibinfo
  {journal} {Phys. Rev.}\ }\textbf {\bibinfo {volume} {D80}},\ \bibinfo {pages}
  {084032} (\bibinfo {year} {2009})},\ \Eprint
  {https://arxiv.org/abs/0906.4983} {arXiv:0906.4983 [gr-qc]} \BibitemShut
  {NoStop}%
\bibitem [{\citenamefont {Ezawa}\ \emph {et~al.}(2010)\citenamefont {Ezawa},
  \citenamefont {Iwasaki}, \citenamefont {Ohkuwa}, \citenamefont {Watanabe},
  \citenamefont {Yamada},\ and\ \citenamefont {Yano}}]{Ezawa2009}%
  \BibitemOpen
  \bibfield  {author} {\bibinfo {author} {\bibfnamefont {Y.}~\bibnamefont
  {Ezawa}}, \bibinfo {author} {\bibfnamefont {H.}~\bibnamefont {Iwasaki}},
  \bibinfo {author} {\bibfnamefont {Y.}~\bibnamefont {Ohkuwa}}, \bibinfo
  {author} {\bibfnamefont {S.}~\bibnamefont {Watanabe}}, \bibinfo {author}
  {\bibfnamefont {N.}~\bibnamefont {Yamada}},\ and\ \bibinfo {author}
  {\bibfnamefont {T.}~\bibnamefont {Yano}},\ }\bibfield  {title} {\bibinfo
  {title} {{On the equivalence theorem in f(R)-type generalized gravity}},\
  }\href {https://doi.org/10.1393/ncb/i2010-10916-1} {\bibfield  {journal}
  {\bibinfo  {journal} {Nuovo Cim.}\ }\textbf {\bibinfo {volume} {B125}},\
  \bibinfo {pages} {1039} (\bibinfo {year} {2010})},\ \Eprint
  {https://arxiv.org/abs/0902.3317} {arXiv:0902.3317 [gr-qc]} \BibitemShut
  {NoStop}%
\bibitem [{\citenamefont {Falls}\ and\ \citenamefont
  {Herrero-Valea}(2019)}]{Falls2018}%
  \BibitemOpen
  \bibfield  {author} {\bibinfo {author} {\bibfnamefont {K.}~\bibnamefont
  {Falls}}\ and\ \bibinfo {author} {\bibfnamefont {M.}~\bibnamefont
  {Herrero-Valea}},\ }\bibfield  {title} {\bibinfo {title} {{Frame
  (In)equivalence in Quantum Field Theory and Cosmology}},\ }\href
  {https://doi.org/10.1140/epjc/s10052-019-7070-3} {\bibfield  {journal}
  {\bibinfo  {journal} {Eur. Phys. J.}\ }\textbf {\bibinfo {volume} {C79}},\
  \bibinfo {pages} {595} (\bibinfo {year} {2019})},\ \Eprint
  {https://arxiv.org/abs/1812.08187} {arXiv:1812.08187 [hep-th]} \BibitemShut
  {NoStop}%
\bibitem [{\citenamefont {Kamenshchik}\ and\ \citenamefont
  {Steinwachs}(2015)}]{Kamenshchik2014}%
  \BibitemOpen
  \bibfield  {author} {\bibinfo {author} {\bibfnamefont {A.~{\relax Yu}.}\
  \bibnamefont {Kamenshchik}}\ and\ \bibinfo {author} {\bibfnamefont {C.~F.}\
  \bibnamefont {Steinwachs}},\ }\bibfield  {title} {\bibinfo {title} {{Question
  of quantum equivalence between Jordan frame and Einstein frame}},\ }\href
  {https://doi.org/10.1103/PhysRevD.91.084033} {\bibfield  {journal} {\bibinfo
  {journal} {Phys. Rev.}\ }\textbf {\bibinfo {volume} {D91}},\ \bibinfo {pages}
  {084033} (\bibinfo {year} {2015})},\ \Eprint
  {https://arxiv.org/abs/1408.5769} {arXiv:1408.5769 [gr-qc]} \BibitemShut
  {NoStop}%
\bibitem [{\citenamefont {Ohta}(2018)}]{Ohta2017}%
  \BibitemOpen
  \bibfield  {author} {\bibinfo {author} {\bibfnamefont {N.}~\bibnamefont
  {Ohta}},\ }\bibfield  {title} {\bibinfo {title} {{Quantum equivalence of
  $f(R)$ gravity and scalar tensor theories in the Jordan and Einstein
  frames}},\ }\href {https://doi.org/10.1093/ptep/pty008} {\bibfield  {journal}
  {\bibinfo  {journal} {PTEP}\ }\textbf {\bibinfo {volume} {2018}},\ \bibinfo
  {pages} {033B02} (\bibinfo {year} {2018})},\ \Eprint
  {https://arxiv.org/abs/1712.05175} {arXiv:1712.05175 [hep-th]} \BibitemShut
  {NoStop}%
\bibitem [{\citenamefont {Benedetti}\ and\ \citenamefont
  {Guarnieri}(2014)}]{Filippo2013}%
  \BibitemOpen
  \bibfield  {author} {\bibinfo {author} {\bibfnamefont {D.}~\bibnamefont
  {Benedetti}}\ and\ \bibinfo {author} {\bibfnamefont {F.}~\bibnamefont
  {Guarnieri}},\ }\bibfield  {title} {\bibinfo {title} {{Brans-Dicke theory in
  the local potential approximation}},\ }\href
  {https://doi.org/10.1088/1367-2630/16/5/053051} {\bibfield  {journal}
  {\bibinfo  {journal} {New J. Phys.}\ }\textbf {\bibinfo {volume} {16}},\
  \bibinfo {pages} {053051} (\bibinfo {year} {2014})},\ \Eprint
  {https://arxiv.org/abs/1311.1081} {arXiv:1311.1081 [hep-th]} \BibitemShut
  {NoStop}%
\bibitem [{\citenamefont {Dirac}(1966)}]{dirac1966}%
  \BibitemOpen
  \bibfield  {author} {\bibinfo {author} {\bibfnamefont {P.~A.~M.}\
  \bibnamefont {Dirac}},\ }\href@noop {} {\emph {\bibinfo {title} {Lectures on
  quantum field theory}}}\ (\bibinfo  {publisher} {Yeshiva Univ.},\ \bibinfo
  {year} {1966})\BibitemShut {NoStop}%
\bibitem [{\citenamefont {Esposito}(1992)}]{Esposito1992}%
  \BibitemOpen
  \bibfield  {author} {\bibinfo {author} {\bibfnamefont {G.}~\bibnamefont
  {Esposito}},\ }\bibfield  {title} {\bibinfo {title} {{Quantum gravity,
  quantum cosmology and Lorentzian geometries}},\ }\href
  {https://doi.org/10.1007/978-3-540-47295-7} {\bibfield  {journal} {\bibinfo
  {journal} {Lect. Notes Phys. Monogr.}\ }\textbf {\bibinfo {volume} {12}},\
  \bibinfo {pages} {1} (\bibinfo {year} {1992})}\BibitemShut {NoStop}%
\bibitem [{\citenamefont {Olmo}\ and\ \citenamefont
  {Sanchis-Alepuz}(2011)}]{Olmo}%
  \BibitemOpen
  \bibfield  {author} {\bibinfo {author} {\bibfnamefont {G.~J.}\ \bibnamefont
  {Olmo}}\ and\ \bibinfo {author} {\bibfnamefont {H.}~\bibnamefont
  {Sanchis-Alepuz}},\ }\bibfield  {title} {\bibinfo {title} {{Hamiltonian
  Formulation of Palatini f(R) theories a la Brans-Dicke}},\ }\href
  {https://doi.org/10.1103/PhysRevD.83.104036} {\bibfield  {journal} {\bibinfo
  {journal} {Phys. Rev.}\ }\textbf {\bibinfo {volume} {D83}},\ \bibinfo {pages}
  {104036} (\bibinfo {year} {2011})},\ \Eprint
  {https://arxiv.org/abs/1101.3403} {arXiv:1101.3403 [gr-qc]}
  \BibitemShut {NoStop}%
\bibitem [{\citenamefont {Gielen}\ \emph {et~al.}(2018)\citenamefont {Gielen},
  \citenamefont {de~Leon~Ardon},\ and\ \citenamefont {Percacci}}]{Gielen}%
  \BibitemOpen
  \bibfield  {author} {\bibinfo {author} {\bibfnamefont {S.}~\bibnamefont
  {Gielen}}, \bibinfo {author} {\bibfnamefont {R.}~\bibnamefont
  {de~Leon~Ardon}},\ and\ \bibinfo {author} {\bibfnamefont {R.}~\bibnamefont
  {Percacci}},\ }\bibfield  {title} {\bibinfo {title} {{Gravity with more or
  less gauging}},\ }\href@noop {} {\bibfield  {journal} {\bibinfo  {journal}
  {Class. Quant. Grav.}\ }\textbf {\bibinfo {volume} {35}},\ \bibinfo {pages}
  {195009} (\bibinfo {year} {2018})},\ \Eprint
  {https://arxiv.org/abs/1805.11626} {arXiv:1805.11626 [gr-qc]} \BibitemShut
  {NoStop}%
\bibitem [{\citenamefont {Floreanini}\ and\ \citenamefont
  {Jackiw}(1987)}]{floreaniniJackiw}%
  \BibitemOpen
  \bibfield  {author} {\bibinfo {author} {\bibfnamefont {R.}~\bibnamefont
  {Floreanini}}\ and\ \bibinfo {author} {\bibfnamefont {R.}~\bibnamefont
  {Jackiw}},\ }\bibfield  {title} {\bibinfo {title} {Self-dual fields as
  charge-density solitons},\ }\href
  {https://link.aps.org/doi/10.1103/PhysRevLett.59.1873} {\bibfield  {journal}
  {\bibinfo  {journal} {Phys. Rev. Lett.}\ }\textbf {\bibinfo {volume} {59}},\
  \bibinfo {pages} {1873} (\bibinfo {year} {1987})}\BibitemShut {NoStop}%
\bibitem [{\citenamefont {Costa}\ and\ \citenamefont
  {Girotti}(1988)}]{costagirotti}%
  \BibitemOpen
  \bibfield  {author} {\bibinfo {author} {\bibfnamefont {M.~E.~V.}\
  \bibnamefont {Costa}}\ and\ \bibinfo {author} {\bibfnamefont {H.~O.}\
  \bibnamefont {Girotti}},\ }\bibfield  {title} {\bibinfo {title} {Comment on
  ``self-dual fields as charge-density solitons''},\ }\href
  {https://link.aps.org/doi/10.1103/PhysRevLett.60.1771} {\bibfield  {journal}
  {\bibinfo  {journal} {Phys. Rev. Lett.}\ }\textbf {\bibinfo {volume} {60}},\
  \bibinfo {pages} {1771} (\bibinfo {year} {1988})}\BibitemShut {NoStop}%
\bibitem [{\citenamefont {Faddeev}\ and\ \citenamefont
  {Jackiw}(1988)}]{faddeevJackiw}%
  \BibitemOpen
  \bibfield  {author} {\bibinfo {author} {\bibfnamefont {L.}~\bibnamefont
  {Faddeev}}\ and\ \bibinfo {author} {\bibfnamefont {R.}~\bibnamefont
  {Jackiw}},\ }\bibfield  {title} {\bibinfo {title} {Hamiltonian reduction of
  unconstrained and constrained systems},\ }\href
  {https://link.aps.org/doi/10.1103/PhysRevLett.60.1692} {\bibfield  {journal}
  {\bibinfo  {journal} {Phys. Rev. Lett.}\ }\textbf {\bibinfo {volume} {60}},\
  \bibinfo {pages} {1692} (\bibinfo {year} {1988})}\BibitemShut {NoStop}%
\bibitem [{\citenamefont {Brans}\ and\ \citenamefont
  {Dicke}(1961)}]{Brans1961}%
  \BibitemOpen
  \bibfield  {author} {\bibinfo {author} {\bibfnamefont {C.}~\bibnamefont
  {Brans}}\ and\ \bibinfo {author} {\bibfnamefont {R.~H.}\ \bibnamefont
  {Dicke}},\ }\bibfield  {title} {\bibinfo {title} {{Mach's principle and a
  relativistic theory of gravitation}},\ }\href
  {https://doi.org/10.1103/PhysRev.124.925} {\bibfield  {journal} {\bibinfo
  {journal} {Phys. Rev.}\ }\textbf {\bibinfo {volume} {124}},\ \bibinfo {pages}
  {925} (\bibinfo {year} {1961})},\ \bibinfo {note} {[,142(1961)]}\BibitemShut
  {NoStop}%
\bibitem [{\citenamefont {Arnowitt}\ \emph {et~al.}(1960)\citenamefont
  {Arnowitt}, \citenamefont {Deser},\ and\ \citenamefont {Misner}}]{ADM}%
  \BibitemOpen
  \bibfield  {author} {\bibinfo {author} {\bibfnamefont {R.}~\bibnamefont
  {Arnowitt}}, \bibinfo {author} {\bibfnamefont {S.}~\bibnamefont {Deser}},\
  and\ \bibinfo {author} {\bibfnamefont {C.~W.}\ \bibnamefont {Misner}},\
  }\bibfield  {title} {\bibinfo {title} {Canonical variables for general
  relativity},\ }\href {https://doi.org/10.1103/PhysRev.117.1595} {\bibfield
  {journal} {\bibinfo  {journal} {Phys. Rev.}\ }\textbf {\bibinfo {volume}
  {117}},\ \bibinfo {pages} {1595} (\bibinfo {year} {1960})}\BibitemShut
  {NoStop}%
\bibitem [{\citenamefont {DeWitt}(1967)}]{DeWitt1967}%
  \BibitemOpen
  \bibfield  {author} {\bibinfo {author} {\bibfnamefont {B.~S.}\ \bibnamefont
  {DeWitt}},\ }\bibfield  {title} {\bibinfo {title} {{Quantum Theory of
  Gravity. 1. The Canonical Theory}},\ }\href
  {https://doi.org/10.1103/PhysRev.160.1113} {\bibfield  {journal} {\bibinfo
  {journal} {Phys. Rev.}\ }\textbf {\bibinfo {volume} {160}},\ \bibinfo {pages}
  {1113} (\bibinfo {year} {1967})}\BibitemShut {NoStop}%
\bibitem [{\citenamefont {Menotti}(2017)}]{Menotti2017}%
  \BibitemOpen
  \bibfield  {author} {\bibinfo {author} {\bibfnamefont {P.}~\bibnamefont
  {Menotti}},\ }\bibfield  {title} {\bibinfo {title} {{Lectures on
  gravitation}}\ }(\bibinfo {year} {2017})\ \Eprint
  {https://arxiv.org/abs/1703.05155} {arXiv:1703.05155 [gr-qc]} \BibitemShut
  {NoStop}%
\bibitem [{\citenamefont {Hojman}\ \emph {et~al.}(1976)\citenamefont {Hojman},
  \citenamefont {Kuchar},\ and\ \citenamefont {Teitelboim}}]{Kuchar1}%
  \BibitemOpen
  \bibfield  {author} {\bibinfo {author} {\bibfnamefont {S.}~\bibnamefont
  {Hojman}}, \bibinfo {author} {\bibfnamefont {K.}~\bibnamefont {Kuchar}},\
  and\ \bibinfo {author} {\bibfnamefont {C.}~\bibnamefont {Teitelboim}},\
  }\bibfield  {title} {\bibinfo {title} {{Geometrodynamics Regained}},\ }\href
  {https://doi.org/10.1016/0003-4916(76)90112-3} {\bibfield  {journal}
  {\bibinfo  {journal} {Annals Phys.}\ }\textbf {\bibinfo {volume} {96}},\
  \bibinfo {pages} {88} (\bibinfo {year} {1976})}\BibitemShut {NoStop}%
\bibitem [{\citenamefont {Kuchar}(1974)}]{Kuchar2}%
  \BibitemOpen
  \bibfield  {author} {\bibinfo {author} {\bibfnamefont {K.}~\bibnamefont
  {Kuchar}},\ }\bibfield  {title} {\bibinfo {title} {{Geometrodynamics regained
  - a Lagrangian approach}},\ }\href {https://doi.org/10.1063/1.1666715}
  {\bibfield  {journal} {\bibinfo  {journal} {J. Math. Phys.}\ }\textbf
  {\bibinfo {volume} {15}},\ \bibinfo {pages} {708} (\bibinfo {year}
  {1974})}\BibitemShut {NoStop}%
\bibitem [{\citenamefont {Garay}\ and\ \citenamefont
  {Garcia-Bellido}(1993)}]{Garay1992}%
  \BibitemOpen
  \bibfield  {author} {\bibinfo {author} {\bibfnamefont {L.~J.}\ \bibnamefont
  {Garay}}\ and\ \bibinfo {author} {\bibfnamefont {J.}~\bibnamefont
  {Garcia-Bellido}},\ }\bibfield  {title} {\bibinfo {title}
  {{Jordan-Brans-Dicke quantum wormholes and Coleman's mechanism}},\ }\href
  {https://doi.org/10.1016/0550-3213(93)90411-H} {\bibfield  {journal}
  {\bibinfo  {journal} {Nucl. Phys. B}\ }\textbf {\bibinfo {volume} {400}},\
  \bibinfo {pages} {416} (\bibinfo {year} {1993})},\ \Eprint
  {https://arxiv.org/abs/gr-qc/9209015} {arXiv:gr-qc/9209015} \BibitemShut
  {NoStop}%
\bibitem [{\citenamefont {Dabrowski}\ \emph {et~al.}(2009)\citenamefont
  {Dabrowski}, \citenamefont {Garecki},\ and\ \citenamefont
  {Blaschke}}]{Dabrowski2008}%
  \BibitemOpen
  \bibfield  {author} {\bibinfo {author} {\bibfnamefont {M.~P.}\ \bibnamefont
  {Dabrowski}}, \bibinfo {author} {\bibfnamefont {J.}~\bibnamefont {Garecki}},\
  and\ \bibinfo {author} {\bibfnamefont {D.~B.}\ \bibnamefont {Blaschke}},\
  }\bibfield  {title} {\bibinfo {title} {{Conformal transformations and
  conformal invariance in gravitation}},\ }\href
  {https://doi.org/10.1002/andp.200810331} {\bibfield  {journal} {\bibinfo
  {journal} {Annalen Phys.}\ }\textbf {\bibinfo {volume} {18}},\ \bibinfo
  {pages} {13} (\bibinfo {year} {2009})},\ \Eprint
  {https://arxiv.org/abs/0806.2683} {arXiv:0806.2683 [gr-qc]} \BibitemShut
  {NoStop}%
\bibitem [{\citenamefont {Kiefer}\ and\ \citenamefont
  {Nikolic}(2017)}]{Kiefer2017}%
  \BibitemOpen
  \bibfield  {author} {\bibinfo {author} {\bibfnamefont {C.}~\bibnamefont
  {Kiefer}}\ and\ \bibinfo {author} {\bibfnamefont {B.}~\bibnamefont
  {Nikolic}},\ }\bibfield  {title} {\bibinfo {title} {{Conformal and
  Weyl-Einstein gravity: Classical geometrodynamics}},\ }\href
  {https://doi.org/10.1103/PhysRevD.95.084018} {\bibfield  {journal} {\bibinfo
  {journal} {Phys. Rev. D}\ }\textbf {\bibinfo {volume} {95}},\ \bibinfo
  {pages} {084018} (\bibinfo {year} {2017})},\ \Eprint
  {https://arxiv.org/abs/1702.04973} {arXiv:1702.04973 [gr-qc]} \BibitemShut
  {NoStop}%
\bibitem [{\citenamefont {Niedermaier}(2020)}]{Niedermaier2020}%
  \BibitemOpen
  \bibfield  {author} {\bibinfo {author} {\bibfnamefont {M.}~\bibnamefont
  {Niedermaier}},\ }\bibfield  {title} {\bibinfo {title} {{Nonstandard Action
  of Diffeomorphisms and Gravity\textquoteright{}s Anti-Newtonian Limit}},\
  }\href {https://doi.org/10.3390/sym12050752} {\bibfield  {journal} {\bibinfo
  {journal} {Symmetry}\ }\textbf {\bibinfo {volume} {12}},\ \bibinfo {pages}
  {752} (\bibinfo {year} {2020})}\BibitemShut {NoStop}%
\bibitem [{\citenamefont {Bonanno}\ \emph {et~al.}(2018)\citenamefont
  {Bonanno}, \citenamefont {Gionti~S.J.},\ and\ \citenamefont
  {Platania}}]{Bonanno2017}%
  \BibitemOpen
  \bibfield  {author} {\bibinfo {author} {\bibfnamefont {A.}~\bibnamefont
  {Bonanno}}, \bibinfo {author} {\bibfnamefont {G.}~\bibnamefont
  {Gionti~S.J.}},\ and\ \bibinfo {author} {\bibfnamefont {A.}~\bibnamefont
  {Platania}},\ }\bibfield  {title} {\bibinfo {title} {{Bouncing and emergent
  cosmologies from Arnowitt Deser Misner RG flows}},\ }\href
  {https://doi.org/10.1088/1361-6382/aaa535} {\bibfield  {journal} {\bibinfo
  {journal} {Class. Quant. Grav.}\ }\textbf {\bibinfo {volume} {35}},\ \bibinfo
  {pages} {065004} (\bibinfo {year} {2018})},\ \Eprint
  {https://arxiv.org/abs/1710.06317} {arXiv:1710.06317 [gr-qc]} \BibitemShut
  {NoStop}%
\bibitem [{\citenamefont {Banerjee}\ and\ \citenamefont
  {Majumder}(2016)}]{Banerjee2016}%
  \BibitemOpen
  \bibfield  {author} {\bibinfo {author} {\bibfnamefont {N.}~\bibnamefont
  {Banerjee}}\ and\ \bibinfo {author} {\bibfnamefont {B.}~\bibnamefont
  {Majumder}},\ }\bibfield  {title} {\bibinfo {title} {{A question mark on the
  equivalence of Einstein and Jordan frames}},\ }\href@noop {} {\bibfield
  {journal} {\bibinfo  {journal} {Phys. Lett.}\ }\textbf {\bibinfo {volume}
  {B754}},\ \bibinfo {pages} {129} (\bibinfo {year} {2016})},\ \Eprint
  {https://arxiv.org/abs/1601.06152} {arXiv:1601.06152 [gr-qc]} \BibitemShut
  {NoStop}%
\end{thebibliography}
\end{document}